\documentclass[9pt, sigconf]{acmart}
\pdfoutput=1
\AtBeginDocument{%
  \providecommand\BibTeX{{%
    \normalfont B\kern-0.5em{\scshape i\kern-0.25em b}\kern-0.8em\TeX}}}

\usepackage[colorinlistoftodos]{todonotes}
\usepackage{algorithmic}
\usepackage{graphicx}
\usepackage{textcomp}
\usepackage{xcolor}
\usepackage{booktabs}
\usepackage{balance}
\usepackage{hyperref}
\usepackage{soul}
\usepackage{enumitem}
\usepackage{natbib}
\usepackage{multirow}
\usepackage{graphicx}  
\usepackage{epstopdf}
\setlist[itemize]{noitemsep, topsep=0pt}

\setcopyright{none}
\renewcommand\footnotetextcopyrightpermission[1]{}
\begin{document}

\title{Studying Vulnerable Code Entities in R}
%
%
%
%
\author{Zixiao Zhao}
\email{zixiaosh@student.ubc.ca}
\orcid{0009-0008-6947-2723}
\affiliation{%
  \institution{University of British Columbia}
  \city{Kelowna}
  \state{BC}
  \country{Canada}
}

\author{Millon Madhur Das}
\email{millonmadhurdas@kgpian.iitgp.ac.in}
\affiliation{%
  \institution{Indian Institute of Technology}
  \city{Kharagpur}
  \state{WB}
  \country{India}
  }

\author{Fatemeh H. Fard}
\email{fatemeh.fard@ubc.ca}
\orcid{1234-5678-9012}
\affiliation{%
  \institution{University of British Columbia}
  \city{Kelowna}
  \state{BC}
  \country{Canada}
}

\begin{abstract}

Pre-trained Code Language Models (Code-PLMs) have shown many advancements and achieved state-of-the-art results for many software engineering tasks in the past few years. These models are mainly targeted for popular programming languages such as Java and Python, leaving out many other ones like R. Though R has a wide community of developers and users, there is little known about the applicability of Code-PLMs for R. 
In this preliminary study, we aim to investigate the vulnerability of Code-PLMs for code entities in R. For this purpose, we use an R dataset of code and comment pairs and then apply CodeAttack, a black-box attack model that uses the structure of code to generate adversarial code samples. 
We investigate how the model can attack different entities in R. 
This is the first step towards understanding the importance of R token types, compared to popular programming languages (e.g., Java). We limit our study to code summarization. 
Our results show that the most vulnerable code entity is the \textit{identifier}, followed by some syntax tokens specific to R.
The results can shed light on the importance of token types and help in developing models for code summarization and method name prediction for the R language. 

\end{abstract}


\maketitle

\section{Introduction}

With the introduction of Transformers and attention mechanisms~\cite{Attention}, pre-trained language models (PLMs) trained on source code --code-PLMs-- have shown advantages in problems related to source code. PLMs such as CodeT5, CodeBERT, and GraphCodeBERT have achieved the state-of-the-art performance in a variety of downstream tasks, such as code summarization~\cite{codet5, unixcoder}, code search~\cite{roberta}, and method name prediction~\cite{GraphCodeBERT}. These models are pre-trained with data-driven pre-training on large-scale code data, making them useful for learning the relationship between code and natural language, and the representation of code. 

While there is a significant volume of ongoing research in the field, the majority of studies on code-PLMs have predominantly focused on extensively used programming languages boasting large user bases, such as C/C++, Python, and Java. There has been a noticeable lack of attention given to other programming languages and low-resource languages (i.e., the languages with limited data), like R. R, designed explicitly for statistical analysis and data-mining applications, stands out as a versatile scripting language with platform independence and rich functionality~\cite{R}. The R language is one of the most popular choices for statistical programming, with a substantial presence indicated by over $20,000$ in open-source R packages on CRAN (Comprehensive R Archive Network) only, and ranked 11th-place in the 2023 edition of the IEEE Spectrum programming language rankings\footnote{IEEE Spectrum programming language rankings\href{https://spectrum.ieee.org/the-top-programming-languages-2023}}.

Building on the rapidly changing developments and increasing attention to PLM, ~\citet{CloserLook} studies the robustness of existing PLMs. 
However, there is a limited exploration of the stability and effectiveness of code-PLMs when applied to low-resource languages. Additionally, there has been a glaring oversight, with no discernible efforts directed toward the investigation of vulnerable code entities or snippets in the R programming language. This neglect highlights the urgent necessity for dedicated research in this domain, given R's prevalence in statistical programming and its escalating importance across diverse applications. Addressing this research gap is paramount for guaranteeing the overall security of software applications developed in R and broadening the horizon of vulnerability detection to encompass a more extensive array of programming languages. Moreover, understanding such information can help develop more robust models for various SE tasks specific to R.

In an effort to address this research gap, we applied CodeAttack~\cite{codeattack}, a black-box attack model that uses the structure of code to generate an adversarial code sample on an R dataset comprising code-comment pairs. We study how different entities can be attacked by the model. We will continue to extend our study by investigating how different programming languages are different from each other when attacked by the model, with an emphasis on learning the identifiers for R to improve the results in the future.

Our results show that:
\begin{itemize}
  \item Although low-resource languages are unseeded during the pre-training phase, Adversarial Attacks are still effective on R.
  \item Similar to popular languages like Java, identifiers are the most important code entities in R.
\end{itemize}

This is the first work that investigates the vulnerability of code entities for R in comparison with other languages. We restrict our study to R and only for code summarization task. 
Our early results are the first step to understanding the working of code-PLMs for R, leading to the development of more robust models specific to this language. We aim to develop such models for code summarization (i.e., comment generation), method name prediction, and code search.
We open source all scripts\footnote{The link to CodeAttack experiments: \url{https://github.com/Sleepyhead01/vulnurable-code-entities-R-analysis}}.

\begin{figure*}[ht]
    \centering
    \includegraphics[width=\textwidth]{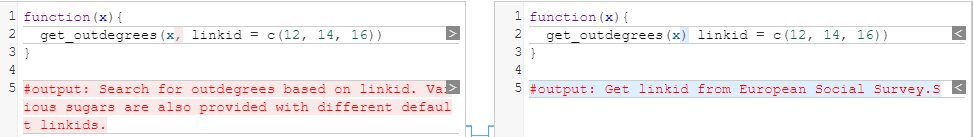}
    \caption{CodeAttack makes a small change to one token in the input code snippet which causes significant changes to the code summary obtained from the SOTA pre-trained programming language models fine-tuned on R.}
    \label{fig:codeattack}
\end{figure*}

\section{Background and Related Work}

\subsection{Code-PLMs and R Studies}

With the introduction of Transformer and the urge of PLMs on natural languages, many Code-PLMs were introduced: CodeBERT~\cite{CodeBERT}, GraphCodeBERT~\cite{GraphCodeBERT}, and CodeT5~\cite{codet5} to name a few. Code-PLMs are mainly trained using CodeSearchNet~\cite{codesearchnet}, which includes Java, Python, PHP, JS, Go, and Ruby, or other datasets that include popular programming languages~\cite{codesearchnet}. CodeXGLUE~\cite{CodeXGLUE} is a main benchmark for code that has a variety tasks to evaluate models that are developed for code. These tasks include Code-Code (e.g., code clone detection or defect prediction), Text-Code (e.g., code generation), or Code-Text (e.g., code summarization). Though there are a lot of models developed~\cite{codet5, CodeBERT, GraphCodeBERT, unixcoder}, to the best of our knowledge, they are not evaluating the results on R. 
Even the StarCoder~\cite{starcoder} that uses R in its pre-training, rarely includes R in its evaluation and does not contain code summarization. The only close work is \cite{CodeCompletionModel} that developed a deep learning model for code completion in R. 
Though there are studies to investigate the dependency of R \cite{DependencyVersioningR, packageDependency} or the R packages \cite{EMSE}, characterizing the bugs in R~\cite{ahmed2023characterizing}, and API breaking changes~\cite{chowdhury2023empirical}, there is no research on R using (code-)PLMs or finding their vulnerability for R.

\subsection{Adversarial Attacks} 
The input-output mappings in deep neural networks are largely discontinuous, and it is possible to add perturbations based on this discontinuity to make the deep learning network predict incorrectly. 
There are several techniques developed to generate adversarial examples using a single gradient step~\cite{FGSM}, calculating the smallest necessary perturbation and applying it to the antagonistic sample construction~\cite{DeepFool}, or utilizing Bayesian optimization to find successful adversarial perturbations with high query efficiency~\cite{BayesOpt}.
Adversarial models are generally divided into white-box and black-box models. 
The white-box models require complete knowledge of the model and the training set. In contrary, the black-box models do not have access to the model parameters, architecture, training data, loss function, or its gradients. The adversary can only query the PLM and get their corresponding output probabilities \cite{codeattack}. 
\subsection{Adversarial Attacks on Code}
ALERT is introduced as a black-box model that attacks the models based on the natural semantics of the generated examples~\cite{ALERT}. The model is applied for three tasks: vulnerability detection, clone prediction, and authorship attribution. However, none of these datasets include R. 
DAMP is another model, which is white-box and changes the variables in code using the gradient information, evaluated on Java and C\#~\cite{10.1145/3428230}. 
ACCENT is introduced in~\cite{10.1145/3501256} and substitutes the identifiers to generate the adversarial examples, and is proposed to improve the robustness of the code comment generation models. Though it uses code summarization, the languages are Java and Python and there is no study on R. 

\textbf{Differences.} Our work is different from these works, as there is no current study of using adversarial models for R, or even code summarization for R. 

\subsubsection{CodeAttack}
CodeAttack is a black-box adversarial attack tool developed for evaluating the robustness of pre-trained models~\cite{codeattack}. CodeAttack represents a form of adversarial attack that aims to deceive models by providing them with misleading input, thereby testing their resilience. Unique to black-box attacks, the attacker lacks access to the model’s internal architecture, parameters, or training data, relying solely on input-output interactions. This scenario is reflective of real-world conditions where attackers often do not possess in-depth knowledge of a system’s internal mechanisms. 
CodeAttack functions by generating adversarial examples designed to lead the model astray, using the CodeBERT model for adversarial code generation.
Figure \ref{fig:codeattack} shows an example of changing one token in the code snippets that leads to a completely different output. These examples are created through minor yet strategic modifications to the input data, inducing the model to produce incorrect outputs. The effectiveness of CodeAttack in compromising model accuracy provides critical insights into the vulnerabilities of these models, thereby informing necessary enhancements in their design and training methodologies.

We choose CodeAttack as this is a state of the arts adversarial model, specifically designed for code. Additionally, it is a black-box approach and is not limited to a specific model. 
CodeAttack can be used to inspect the tokens that the code-PLM attends to, thus giving us insights about important and vulnerable R entities. Additionally, code Attack can only query the code-PLM with input sequences and get their corresponding output probabilities, without requiring access to its parameters. It works irrespective of the programming language, making it a suitable model to use in our study.

\section{Methodology}



We aim to understand the vulnerable tokens in R, meaning the tokens that a small change in them, will affect the performance of the model significantly. 
There are two primary steps: (i) Fine-tune an existing code-PLM on the R dataset, and (ii) apply CodeAttack to the fine-tuned model to obtain the result.

\subsection{Fine-tune a Code-PLM on R}
\subsubsection{CodeT5}
The pre-trained victim model chosen for this study is CodeT5~\cite{codet5}. In the empirical study proposed by \citet{EmpiricalComparison}, most of these
models rely on either using only Transformer-Encoder (TE, the encoder-only)~\cite{CodeBERT, GraphCodeBERT, Cubert, unixcoder}, or use Transformer-Decoder only, like GPT~\cite{GPT}. This is sub-optimal for generation and understanding tasks, respectively. 


CodeT5~\cite{codet5} intends to use the full Transformer architecture to capture the large semantic and structural information from the large dataset it is pre-trained on. 
CodeT5 leverages user-written code annotations and a bimodal generation job to improve Natural Language (NL)-Programming Language (PL) alignment. Tests show that CodeT5 significantly outperforms previous methods in understanding generation tasks in all directions (including PL-NL, NL-PL, and PL-PL~\cite{codet5}), as well as tasks such as code error and clone detection.
Therefore, we chose CodeT5 as the code-PLM; worth noting, CodeT5 is used in the CodeAttack study.


\subsubsection{R dataset}
The dataset we used in this study contains 32,671 R functions collected from GitHub open-source repositories. This dataset is counted in a form similar to the CodeSearchNet~\cite{codesearchnet} dataset, where each piece of data is in the form of a code-comment pair, where the code part has the body of the function, and the comment is a natural language description of what the function does.
This dataset was collected initially for an empirical study on R using code-PLMs. Roxygen, a popular documentation system for R similar
to JavaDoc, is used to parse the functions into abstract syntax trees and obtain the comments. Due to the scripting nature of R, Roxygen is used to ensure code comments can be extracted. The dataset is collected from 1664 GitHub repositories, with the criteria being a public repository, the project being an R package, and the project is active and up to date, excluding the ones that are personal projects or archived. 

\section{Experiment and result}
\subsection{Experiment Setup}
To facilitate a meaningful comparison, we aim to recreate the training environment to closely match the reported. CodeAttack requires a code-PLM trained on a specific task. 
As no code-PLM is trained on R and for code summarization, we cannot directly use the current models. So, we should first fine-tune the CodeT5 model on the R dataset for code summarization. We modified the CodeT5 GitHub repository to work with the R dataset. 
Due to hardware limitations, the GPU memory we had access to does not support the batch size mentioned in CodeT5~\cite{codet5}. Hence, We conducted experiments using batch sizes of 48 and 32 across all the models outlined in Table~\ref{tab:t5}. The reported results are based on the superior BLEU score (BLEU is a score used to evaluate code summarization) obtained between the two batch sizes. It is worth mentioning that for a batch size of 48, distributed training was employed to address constraints related to low GPU memory.
In Table~\ref{tab:t5}, \textit{CodeT5 Reported} is the original results reported in the original paper~\cite{codet5}, \textit{CodeT5 ReTrain} is the model we retrained, \textit{CodeT5-R-base} is the model fine-tuned on the R dataset with the batch size of 48 and \textit{CodeT5-R-small} is the model fine-tuned on the R dataset with the batch size of 32. The results of the replication are generally consistent with the reported results.
As the BLEU score is better for batch size 32, we used this version in our following experiments. 

\begin{table}[]
  \centering
  
  \begin{tabular}{lcc}
    \toprule
    Model & Batch Size & Bleu Score \\
    \midrule
    CodeT5 Reported & 48 & 19.55 (Average across all PL) \\
    CodeT5 ReTrain & 48 & 19.47 (Average across all PL) \\
    CodeT5-R-base & 48 & 15.96 \\
    CodeT5-R-small & 32 & 17.73 \\
    \bottomrule
  \end{tabular}
  \caption{BLEU score of CodeT5 finetuned on R dataset for code summarization.}
  \label{tab:t5}
\end{table}

\subsection{CodeAttack Result}



CodeAttack uses a generator and an adversarial model. In our case, we used the fine-tuned CodeT5~\cite{codet5} on R as the generator model and CodeBERT~\cite{CodeBERT} as the adversarial model.
Please note that CodeT5~\cite{codet5} and CodeBERT~\cite{CodeBERT} are used in CodeAttack, and code summarization was also one of the considered tasks in the original paper. Therefore, we first reproduced the results using the three languages studied originally and then applied them to R.

CodeAttack's goal is to degrade the quality of the generated output sequence while making minimal perturbations in input codes. 
It first requires identifying the vulnerable tokens, which are the tokens having a high influence on the output logits of the model (in the black-box attack we have access to the output logits of the model). The hypothesis of CodeAttack is that keywords and identifiers get more attention values in the CodeBERT model.
CodeAttack has two phases for attacking the victim models: i) finding the most vulnerable tokens, and ii) substituting these vulnerable tokens with adversarial samples. To find vulnerable input tokens, they mask a token in an input sequence, feed this sequence
to the model, and retrieve output logits for this input. Then the influence of the masked token is computed by subtracting the logits of the input containing the masked token from the initial logits of
non-adversarial input. 
Then for more tokens, they rank the tokens by their influence and select the top-k tokens as the most vulnerable ones.

To substitute vulnerable tokens, they use a greedy search method subject to code-specific constraints. In the previous step, the top-k vulnerable tokens in an input sequence are obtained. Next, these vulnerable tokens are changed to turn the input sequence into an adversarial sample. To do so, first, the input sequence is fed to a masked code-PLM, in order to retrieve top-k predictions for the masked tokens in the input sequence. These top-k predictions are the initial search space. The model uses the WordPiece algorithm for tokenization, so they align and mask all sub-words for vulnerable tokens and then combine the predictions to find top-k substitutes for the vulnerable token.
These substitutes are filtered to ensure minimal perturbation, code consistency, and code fluency. The purpose of the code specific constraints is to ensure that the generated tokens are meaningful code tokens.

To use CodeAttack, we referred to their repository. There are specific hyperparameters, including $BLEU\theta$ but there is limited information about this parameter. Therefore, we experimented with various settings to obtain the results. 
Table \ref{tab: result} shows the results. 
$\Delta\textsubscript{drop}$ means the drop of performance of the downstream task, before and after the attack, in our case, the drop in BLEU score.
Interestingly, the drop in the performance of the model for PHP, Python, and Java is higher than the performance drop in R. This could be related to setting the $BLEU\theta$ parameter, or to the dataset. Finding the reasons for this difference requires more experiments and research that is out of the scope of this work. 

\begin{table}[]

\begin{tabular}{llllll}
\hline
PL        & $BLEU\theta$ & $BLEU\textsubscript{before}$    & $BLEU\textsubscript{after}$ & $\Delta\textsubscript{drop}$ \\ \hline
\multirow{3}{*}{R} & 50         & \multirow{3}{*}{16.97} & 11.88               & 5.09       
\multirow{3}{*}{} \\
                   & 60         &                        & 12.19               & 4.78              \\
                   & 75         &                        & 12.71               & 4.25        \\ \hline
PHP                & unreported    & 20.06                  & 11.06               & 9.59                    \\
Python             & unreported    & 20.36                  & 7.97                & 12.39                   \\
Java               & unreported    & 19.77                  & 11.21               & 8.56                    \\ \hline
\end{tabular}
\caption{Results for R, and reproduced results for PHP, Python, and Java on code summarization. The performance is measured in BLEU.}
\label{tab: result}
\end{table}

\subsection{Discussions}

To gain insights into the reported results and satisfy the goals of our study, we investigate the success rate of different token types (i.e., code entities), to find out which ones are the most vulnerable ones. 
First, we used tree-sitter to identify the token types in R, using the R grammar from r-lib\footnote{\url{https://github.com/r-lib/tree-sitter-r}}. 
Then, we calculate the success rate (i.e., success \%) for the token types. Success rate is defined as the percentage of successful attacks (which is measured in $\Delta\textsubscript{drop}$). The higher value means that the adversarial attack was more effective.

\begin{figure}[]
    \centering
    \includegraphics[width=1.0\linewidth]{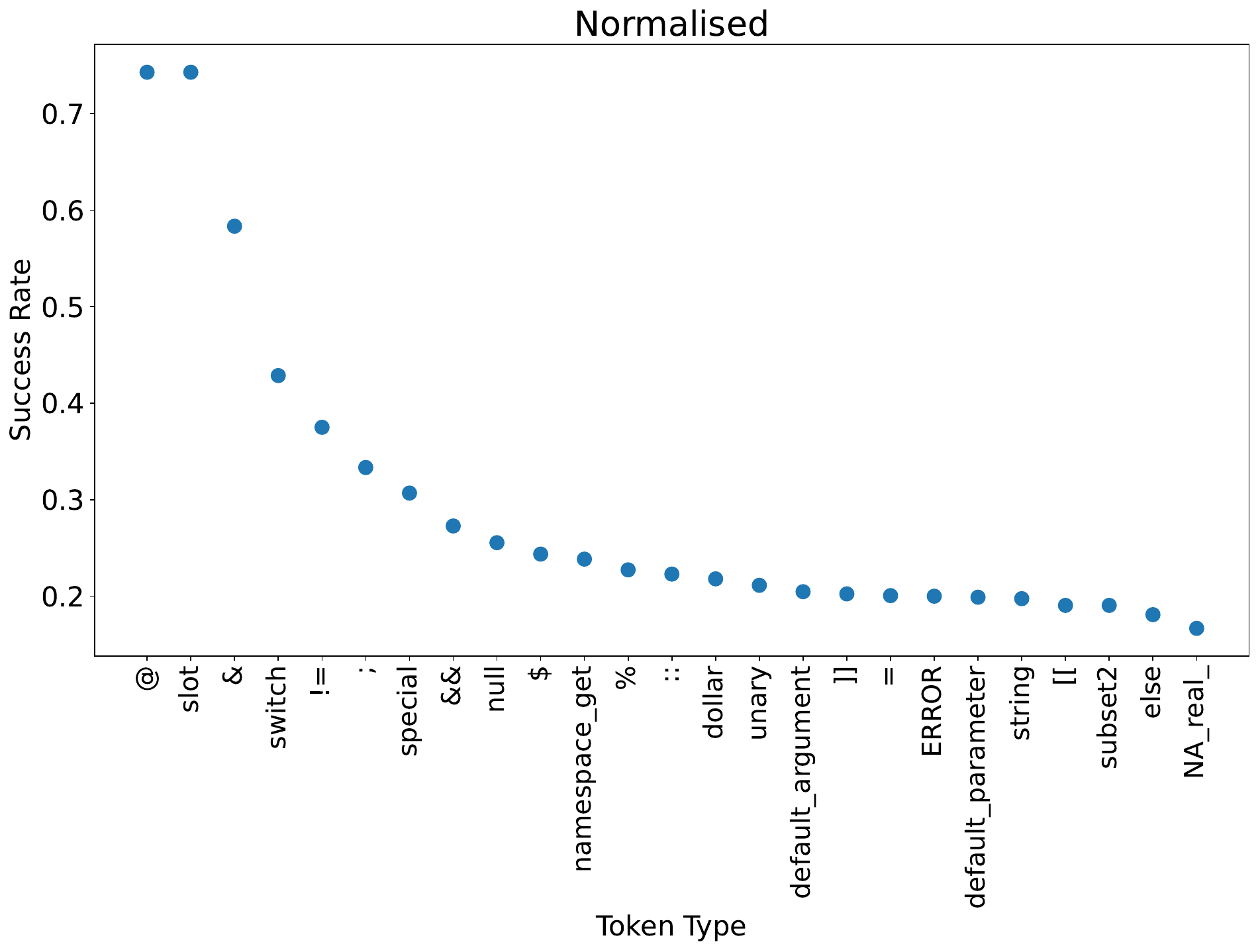}
    \caption{Normalized success rate of different token types in R.}
    \label{fig:token-type-success-rate}
\end{figure}

The results are shown in Figure~\ref{fig:token-type-success-rate}. The normalized plot shows the success rates for different token types. 
The figure shows the normalized plot of the success rate for each token type. 
The normalization is done as, for a token type across all examples, the `number of successful attacks on a specific token' is divided by the `number of times the token is present in the dataset'. 

\begin{figure}[]
    \centering
    \includegraphics[width=1.0\linewidth]{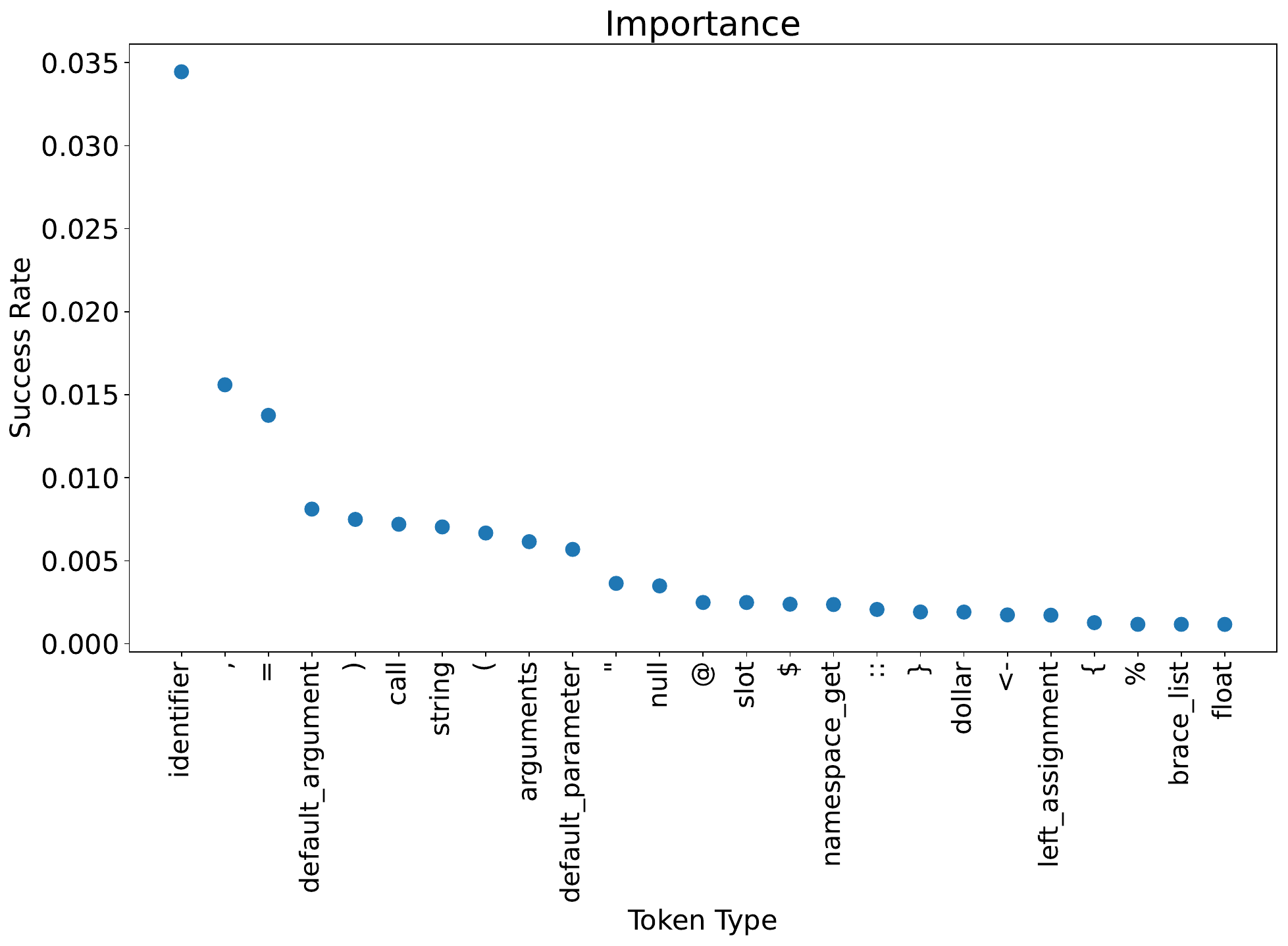}
    \caption{The importance of the code entities (i.e., token types). }
    \label{fig:importance}
\end{figure}

The importance of the token types is shown in Figure~\ref{fig:importance}. The importance is calculated as the `normalized success rate of a token type' divided by the `number of times that specific token was attacked'. 
This analysis shows which token type is the most important token to attack.
The higher this number is, the more successful attacks we will have for that token type. 

Due to space limitations, we restrict the figure size and avoid adding it as a teaser figure. We also restricted the number of tokens to show in Figures~\ref{fig:token-type-success-rate} and \ref{fig:importance}. The results show that the most important token type is the `identifier'. 
This supports the fact that identifiers are known as important tokens in code-related tasks and language models \cite{codet5, MultilingualTraining}. 
Interestingly, the other token types are \verb|`|, \verb|=|, \verb|default-argument|, and \verb|)|. Although backtick (i.e., \verb|`|) is uncommon in other languages, it plays an important role in R. Backticks are used to refer to identifiers that are reserved or illegal in R, known as the non-syntactic names. To refer to such variables, developers need to surround them with backticks~\cite{RforDataScience}. As in other languages, the equal sign, \verb|=|, also serves as a variable declaration and assignment in R, which makes it vulnerable since replacing the equal sign will affect the deconstruction of the program. Based on the result, the most vulnerable tokens are tokens that will provide semantic and structural information about themselves and other tokens, other tokens' semantics may act in a way that defeats its purpose without them.
This might show that the structural information of the R code is also important to be captured, as identified in previous works for other languages \cite{codeattack}.

\section{THREATS TO VALIDITY}

\textbf{Internal:} 
One threat to the obtained results is related to the collected dataset. In R, there are a series of packages collectively referred to as the Tidy verse. Tidyverse allows users to start directly from the manipulation of data, allowing beginners to learn data processing and visualization applications in the shortest possible time. Tidyverse can do this because it has a syntax structure that is very different from that of the R language. This syntax gap can influence the results, as both Tidyverse and base R code samples were combined in the dataset.
Another concern regarding the dataset is that due to the nature of R and Roxygen, some functions in the dataset have the same comments. This might affect the results, though we anticipate it to be low and require further investigation. 

\textbf{External:} 
As noted previously, the repository of CodeAttack is not well documented, and some details were not included. Therefore, we experimented with different settings for the parameters and reproduced the results to alleviate this threat. 
However, as the value of $BLEU\theta$ plays an important role in the results, there might be some threats associated with it. 
Another possible threat is related to the r-lib that we used to parse the R code. To alleviate possible threats, we checked the results manually for some examples and ensured about their correctness. 

\section{Conclusion}
In this preliminary study, we investigated the vulnerability of various code entities in R. Our results show that the importance of identifiers and some syntactical tokens are higher than the rest, suggesting the potential of using the code structure in the training of code-PLMs for this language. This was the first step towards our goal of developing a model specific to R, for tasks such as code summarization and method name prediction. The current work was used to find out the importance of different code tokens in R, which can be valuable in developing more robust models.
\begin{acks}
This research is support by a grant from the Natural Sciences and Engineering Research Council of Canada RGPIN-2019-05175 and Mitacs Accelerate award, 2021 and Mitacs Globalink award, 2023.\textbf{}
\end{acks}

\balance
\bibliographystyle{ACM-Reference-Format}
\bibliography{references}

\end{document}